\documentclass[]{article}

\usepackage{amssymb}
\usepackage{graphicx}\usepackage{amsmath,amssymb}
\usepackage{color}
\usepackage[english]{babel}
\usepackage{xcolor}
\usepackage{srcltx}
\usepackage{greekbf}
\usepackage{nicefrac}
\usepackage{empheq}
\usepackage{enumerate}
\usepackage{wrapfig}
\usepackage{amsthm}

\usepackage{multicol}

\usepackage{mathtools}
\mathtoolsset{showonlyrefs}
    \mathtoolsset{centercolon}
\usepackage{mathrsfs}
\usepackage{epstopdf}

\allowdisplaybreaks

\DeclareMathAlphabet{\mathbf}{OT1}{cmr}{bx}{it}
\DeclareMathAlphabet{\mathssb}{OT1}{cmss}{bx}{n}
\DeclareMathAlphabet{\mathssn}{OT1}{cmss}{m}{n}
\DeclareMathAlphabet{\mathub}{OT1}{cmr}{b}{n}

\DeclareMathAlphabet{\mathpzc}{OT1}{pzc}%
                                 {m}{it}

\theoremstyle{definition}
\newtheorem{remark}{Remark}
\newcommand{\bydef}{\,\raise.050ex\hbox{\rm:}\kern-.025em\hbox{\rm=}\,}
\newcommand{\defby}{=\raise.075ex\hbox{\kern-.325em\hbox{\rm:}}\,}

\def\qed{\relax\ifmmode\hskip2em \Box\else\unskip\nobreak\hskip1em $\Box$\fi}
\newcommand {\eps} {\varepsilon} 
\newcommand {\0} {\textbf{0}}    
\newcommand {\Gc}  {\mathcal{G}}

\newcommand {\Pc}  {\mathcal{P}}

\newcommand {\Sc}  {\mathcal{S}}

\newcommand {\ab} {\mathbf{a}}

\newcommand {\cb} {\mathbf{c}}
\newcommand {\db} {\mathbf{d}}
\newcommand {\eb} {\mathbf{e}}
\newcommand {\fb} {\mathbf{f}}
\newcommand {\gb} {\mathbf{g}}

\newcommand {\lb} {\mathbf{l}}
\newcommand {\mb} {\mathbf{m}}
\newcommand {\nb} {\mathbf{n}}
\newcommand {\pb} {\mathbf{p}}
\newcommand {\qb} {\mathbf{q}}
\newcommand {\rb} {\mathbf{r}}

\newcommand {\tb} {\mathbf{t}}

\newcommand {\ub} {\mathbf{u}}
\newcommand {\vb} {\mathbf{v}}

\newcommand {\Ab} {\mathbf{A}}
\newcommand {\Bb} {\mathbf{B}}
\newcommand {\Cb} {\mathbf{C}}

\newcommand {\Eb} {\mathbf{E}}
\newcommand {\Fb} {\mathbf{F}}

\newcommand {\Ib} {\mathbf{I}}

\newcommand {\Mb} {\mathbf{M}}

\newcommand {\Qb} {\mathbf{Q}}

\newcommand {\Sb} {\mathbf{S}}

\newcommand {\Vb} {\mathbf{V}}
\newcommand {\Wb} {\mathbf{W}}

\newcommand {\Co} {\mathbb{C}}

\newcommand {\Qo} {\mathbb{Q}}

\font\mbo=cmmib10 scaled \magstephalf

\newcommand{\taub}   {\hbox{\mbo {\char 28}}}

\newcommand {\chib}      {\mathbf{\chi}}

\def\eps{\varepsilon}
\newcommand{\sym}{\mathop{\mathrm{sym}}}

\begin{document}

\begin{center}
 {\bf \Large
A Shell Theory for Chiral Single-Wall\\ Carbon Nanotubes}
\end{center}
\medskip

\begin{center}
 {\large
Antonino Favata$^\star$, 
Paolo Podio-Guidugli$^\diamond$}
\end{center}

\begin{center}
 {
$^\star$Institute of Continuum Mechanics and Material Mechanics,\\ Hamburg University of Technology\footnote{Ei\ss endorfer Stra\ss e 42, 21073 Hamburg Germany. Email: antonino.favata@tuhh.de.}
}
\end{center}

\begin{center}
 {
$^\diamond$Dipartimento di Ingegneria Civile e Ingegneria Informatica,\\ Universit\`a di Roma TorVergata\footnote{Via del  Politecnico 1, 00133 Rome, Italy. Email: ppg@uniroma2.it.}
}
\end{center}

\begin{abstract}
In this paper, we propose a characterization of the mechanical response of the linearly elastic shell we associate to a single-wall carbon nanotube of arbitrary chirality. In \cite{Fa2}, we gave such a characterization in the case of zigzag and armchair nanotubes; in particular, we showed that the \emph{orthotropic} response we postulated for the associated shells is to become \emph{isotropic} in the graphene-limit, that is, when the shell radius grows bigger and bigger.  
Here we give an explicit recipe to construct the generally \emph{anisotropic} response of the shell associated to a nanotube of any chirality in terms of the response of the shell associated to a related zigzag or armchair nanotube. The expected coupling of mechanical effects that anisotropy entrains is demonstrated in the case of a \emph{torsion problem}, where the axial extension accompanying twist is  determined analytically and found in good agreement with the available experimental data.
\end{abstract}

\section{Introduction}
In this paper, we deal with  \emph{single-wall carbon nanotubes}, for which we use the abbreviated acronym CNTs, \emph{of arbitrary chirality}.  Given the centrality of the chirality concept in our developments to come, we find it appropriate to begin by a short account of chirality-related concepts, to be safely skipped by a conversant reader.
\subsection{Geometrical premiss}
\begin{figure*}
\centering
\includegraphics[scale=0.9]{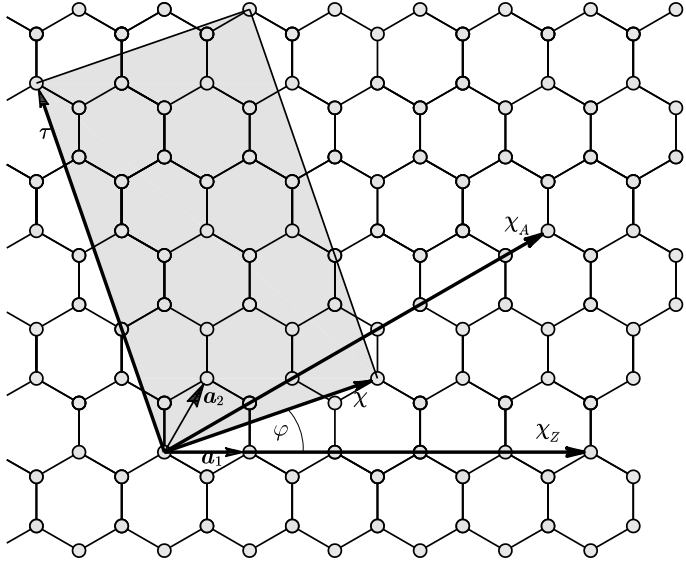}
\caption{The graphene part involved in rolling up a  (2,1)-chiral CNT.} \label{chiral}
\end{figure*}

In imagination, a CNT can
be obtained by rolling up into a cylindrical shape a
\emph{graphene} -- that is, a monolayer flat sheet of graphite -- visualized as a two-dimensional lattice with hexagonal unit cell.  There are many ways to roll a
graphene up, sorted by introducing a geometrical object, the
\emph{chiral vector}:
\begin{equation}\label{chiv}
\chib=n\ab_1+m\ab_2,\quad n\geq m,
\end{equation}
where $n,m$ are two integers, and $\ab_1, \ab_2$ are two \emph{lattice vectors}, such as those at a mutual angle of $\pi/3$ radians shown in Fig. \ref{chiral}. 
Once the lattice vectors are fixed, the ordered pair $(n,m)$ specifies the \emph{chirality} of the CNT to be, whose axis and minimal length are specified by the \emph{axial vector}
\begin{equation}
\taub=t_1\ab_1+t_2\ab_2,\quad 
\end{equation}
where $t_1$ and $t_2$ are integers such that
\begin{equation}
t_1=\frac{n+2m}{d_R}, \quad t_2=-\frac{2n+m}{d_R}, 
\end{equation}
with
$$
d_R:=\gcd(2n+m,n+2m).
$$
It is not difficult to check that $\chib\cdot\taub=0$. 
Likewise, it is the matter of a straightforward calculation to derive the following formula for the \emph{chiral angle} $\varphi=:\displaystyle{\arccos\frac{\chib\cdot\ab_1}{|\chib|\,|\ab_1|}}$\,: 
\begin{equation}\label{chirang}
\tan\varphi=\sqrt3\,\frac{m}{2n+m}\,.
\end{equation} 
When $n>m$, the CNT under examination is termed \emph{chiral}. The $(n,0)$- and
$(n,n)$-nanotubes, at times referred to collectively as \emph{achiral}, are termed, respectively, \emph{zigzag} and
\emph{armchair}; in Fig. \ref{chiral}, their chiral vectors are denoted by, respectively, $\chib_Z\equiv\ab_1$ ($\varphi_Z=0$ radians) and $\chib_A$ ($\varphi_A=\pi/6$ radians).\footnote{Needless to say, a regular hexagonal lattice has the symmetries of an equilateral triangle at each of its points. Therefore, any rotation of an integer multiple of $\pi/3$ maps $(\ab_1,\ab_2)$ into an equivalent pair of lattice vectors.}
The  \emph{nominal radius} $\rho_0$ of a $(n,m)$-CNT is defined to be the radius of the cylinder on which the centers of the C atoms are placed after an ideal rolling-up operation entailing no energy expenditure for the inevitable distortion of the C-C bonds; according to this definition, 
\begin{equation}\label{geonec}
\rho_0=\widehat\rho_0(n,m):=\frac{\sqrt 3}{2\pi}\,n\sqrt{1+m/n+(m/n)^2}\,s,
\end{equation}
where $s$ is the length of the C-C bond;\footnote{In \cite{BFPPG}, $\rho_0$ was called the \emph{geometrically necessary} radius. In terms of a stick-and-spring model of discrete structure mechanics for graphene, where an axial spring aligned with a stick opposes stretching of the corresponding C-C bond and a spiral spring between two sticks opposes changes in their angle, the (geometrically necessary $\equiv$) nominal radius is the radius of the cylinder obtained, in imagination, by rolling graphene up after disconnecting all springs. That elastic energy is stored in a real CNT is demonstrated by the `unzipping' experiments reported in \cite{unz}.} we note here for later reference the following consequence of $\eqref{chiv}_2$:
\[
1\leq \sqrt{1+m/n+(m/n)^2}\leq \sqrt{3}.
\]

\subsection{Introductory remarks}
Chirality is a geometrical character that influences heavily the mechanical, electrical, and thermal, properties of a CNT, especially when its radius is small;  in particular, a variety of chirality-dependent mechanical phenomenologies is described in  \cite{Cha,CC,CGG,LU,WZLJ,WYKB}.  
In this paper, we concentrate on the \emph{influence of chirality on the mechanical response of CNTs}, when they are \emph{modeled as linearly elastic shells}; we propose and use hereafter for this type of model the acronym CNS, standing for \emph{Carbon NanoShell}. 

Searching the literature, one finds many theoretical studies aimed to capture the mechanics of chiral CNTs. For example, within the framework of discrete structure mechanics, stick-and-spiral models  are used in \cite{CG} and \cite{SL} to determine how the elastic properties of chiral CNTs depend on size; closed expressions for chirality and size dependences of elastic properties of CNTs are given in \cite{CG,CGG1}, on the basis of a discrete model; and, in \cite{Cha}, the ideas of \cite{CG,CGG1} are employed with a view to adapt Donnell's theory of linearly elastic shells to chiral CNTs. Moreover, this time in the context of continuum structure mechanics, chirality-dependent properties have been investigated in \cite{Ru}, where a CNS model is proposed for CNTs of arbitrary chiral angle, whose linearly elastic response is deduced from the  orthotropic \textit{plane-stress} response by a procedure involving a small-angle rotation of the coordinate system.

In \cite{Fa2}, we constructed a mechanical model of linearly elastic, \emph{orthotropic} shell, and solved explicitly the relative equilibrium equations in terms of displacements for the cases of axial traction, torsion, inner pressure, and rim flexure. That shell model  depends on a list of seven parameters: two are geometric, thickness and radius, and five constitutive, four of which are independent. In \cite{BFPPG}, our goal was to apply the theory developed in \cite{Fa2} to CNTs, so as to obtain a theory of CNSs. To do so, in the first place we gave precise definitions for the geometric parameters, neither of which has a self-evident one;\footnote{Think, in particular, of the scattered evaluations of an effective wall thickness that led to the formulation of the so-called \emph{Yacobson paradox} \cite{Sh}.} then, as others did before although in a different manner, we determined all parameters in terms of the two nanoscale constants measuring the extensional and dihedral energies of C-C bonds. The resulting \emph{nanoscopically informed} CNS theory is applicable to both \emph{zig-zag and armchair CNTs}, its predictions matching experiments fairly well. 

We here generalize the theory developed in \cite{Fa2} and  \cite{BFPPG} so as to obtain a theory of  \emph{anisotropic} linearly elastic shells that reproduces fairly well the mechanical behavior of a single-wall CNT of \emph{arbitrary chirality}. A relevant feature of our theory is that a number of equilibrium problems formulated within it can be shown to have \emph{explicit analytic solutions}. An example we work out in detail exhibits the expected \emph{coupling of torsional and extensional effects}, in good agreement with the available experimental data. 

\subsection{Summary of contents}
Firstly, we lay out the constitutive assumptions of our shell theory. Precisely, in Section \ref{AD} we specify that the admissible displacements are of the Kirchhoff-Love type; in Section \ref{AC}, we introduce a consistent representation for the elasticity tensor of a $(n,0)$-zigzag CNT; and, in Section \ref{C}, we give an argument to arrive at a simple formula that yields the elasticity tensor of the CNS associated to a given $(n,m)$-CNT in terms of the associated $(n,0)$-zigzag CNT, via an operation of \emph{orthogonal conjugation} that depends in an explicit form on the chirality parameters $n$ and $m$. With this, the dependence on chirality of all the constitutive parameters that enter our theory is completely specified, and we pass to the geometrical ones.

While we do not reproduce here from  \cite{BFPPG} the lengthy reasoning that led us to propose precise definitions for both \emph{effective thickness} and \emph{effective radius} of the CNS to be associated to a given $(n,n)$-armchair or $(n,0)$-zigzag CNT, in Section \ref{RT} we do reproduce from that paper  two curves allowing for a visualization of the dependence of those parameters on the chirality index $n$. As those curves make evident, the small differences in effective thickness existing for $n$ small do disappear very quickly; likewise, both for armchair and zigzag CNTS, the ratio of effective-to-nominal radii tends rather quickly to 1 when $n$ grows. Consequently, for the CNS associated to a given a $(n,m)$-CNT we take the effective thickness of a $(n,0)$-CNT, and we take the effective radius equal to the nominal radius given by \eqref{geonec}. 

The point-wise balance equations and boundary conditions of our CNS theory are derived in Section 4.2, from a two-dimensional Principle of Virtual Power that we deduce in Section 4.1 from a suitable three-dimensional PVP by a procedure detailed in  \cite{BFPPG}. Section 4.3 deals with the special and simpler axisymmetric problems, among which is the torsion problem when two balancing torques are applied at a CNS's ends. As Fig.s \ref{torsang}, \ref{rigtor} and \ref{axstr} show, given the torques and fixed the chiral number $n$, the torsion angle diminishes, the torsion stiffness grows bigger, and the axial strain as a centered maximum, when $m$ grows from 1 to $n$. 

\section{Constitutive Assumptions}
We let the shell-like body of interest be a tubular neighborhood $\Gc(\Sc, \eps)$, of constant thickness $2\eps$, of
a right circular cylinder $\Sc$ of radius $\rho_o$; we defer to Section \ref{RT} a discussion of how these geometric parameters depend on chirality. Following \cite{Fa2}, we define the mechanical response of $\Gc(\Sc, \eps)$ by selecting a class of \emph{admissible deformations} and by making a consistent choice for the \emph{elasticity tensor} $\Co$ of the material $\Gc(\Sc, \eps)$ is comprised of. 

\subsection{Admissible deformations}\label{AD}
\begin{figure}[h]
\centering
\includegraphics[scale=0.7]{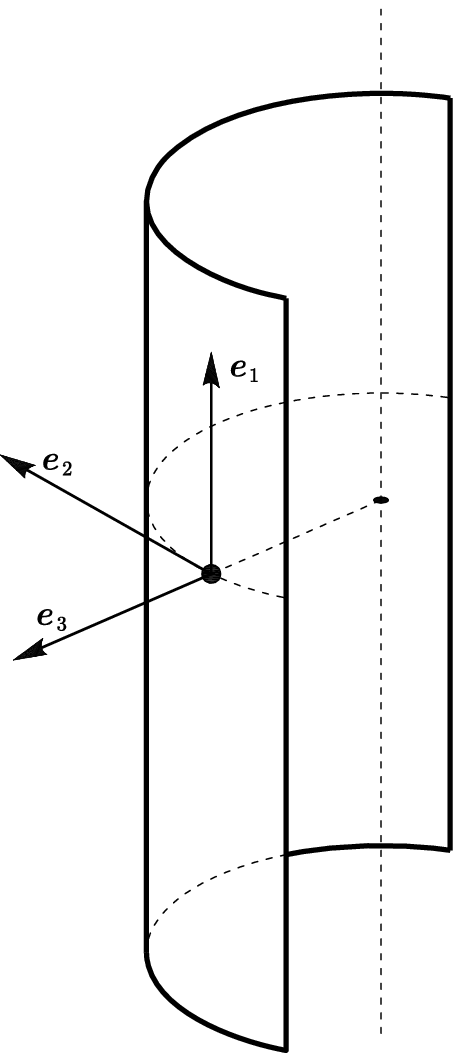}
\caption{} \label{assi_fig}
\end{figure}

At any fixed point of $\Gc(\Sc, \eps)$, let $\{\eb_i\,|\,i=1,2,3\}$ be the orthonormal vector basis shown in Fig. \ref{assi_fig}.

\noindent For the linear space of all symmetric tensors, we find it convenient to use the following basis:
\begin{equation}\label{base}
\begin{aligned}
\Vb_\alpha&=\frac{1}{\sqrt{2}}(\eb_\alpha\otimes\eb_3+\eb_3\otimes\eb_\alpha)\;\,(\alpha=1,2),\\ \Vb_3&=\eb_3\otimes\eb_3,\\
\Wb_\alpha&=\eb_\alpha\otimes\eb_\alpha \;\,
(\alpha\,\;\textrm{not summed}),\\
\Wb_3&=\frac{1}{\sqrt{2}}\,(\eb_1\otimes\eb_2+\eb_2\otimes\eb_1).
\end{aligned}
\end{equation}

The admissible deformations  of $\Gc(\Sc, \eps)$, whatever the chirality of the CNT whose mechanical response $\Gc(\Sc, \eps)$ is intended to model, are those induced by displacement fields  $\ub$ such that the associated strains 
\begin{equation}\label{strain}
\Eb(\ub)=\sym \nabla \ub:=\frac{1}{2}(\nabla\ub+\nabla\ub^T)
\end{equation}
satisfy the \emph{Kirchhoff-Love constraint}
\begin{equation}\label{KL}
\Eb(\ub)\eb_3=\0,
\end{equation}
or rather, equivalently, have the following form in the tensor basis \eqref{base}:
\begin{equation}\label{bonstrain}
\Eb(\ub)=e_{i}(\ub)\Wb_i\,.
\end{equation}

In \cite{Fa2}, other less stringent internal constraints are considered, for shell models of multi-wall CNTs. Here and in \cite{BFPPG}, attention is restricted to single-wall CNTs, for which the meager kinematics of Kirchhoff--Love's theory suffices.
A family of axisymmetric equilibrium problems that admit a Kirchhoff-Love solution $\ub$ will be formulated in Section 4; and, in Section 5, the \emph{torsion problem} will be solved explicitly, for an arbitrary choice of chirality.

\subsection{The response of achiral CNSs}\label{AC}
The symmetries of a regular hexagonal lattice are those of an equilateral triangle. Consequently, graphene, when regarded as a flat continuous body to be deformed exclusively in its own plane, is assigned an \emph{isotropic} linearly elastic response (see e.g. \cite{AR} and \cite{CD}). However, rolling-up destroys the local symmetries that guarantee isotropy in the flat case. Therefore, what response symmetries to assign to a CNT modeled as an elastic shell becomes an issue. 

\begin{figure*}
\centering
\includegraphics[scale=0.8]{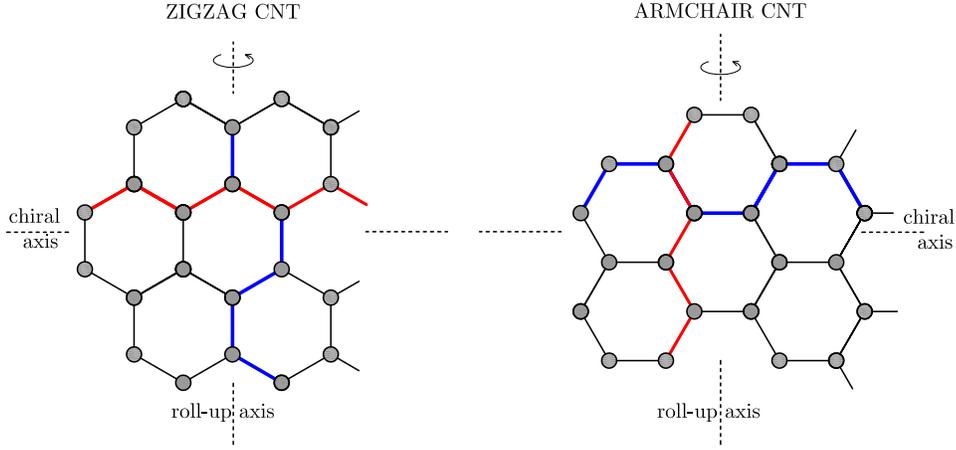}
\caption{Roll-up and chiral axes of zigzag and armchair carbon nanotubes. Note the
orthogonality of zigzag (red) and armchair (blue) atom sequences.} \label{rollup}
\end{figure*}

As a glance to Fig. \ref{rollup} makes evident, 
both for zigzag and for armchair CNTs there are, at each lattice point in the graphene plane,  two orthogonal directions that remain orthogonal after a rolling-up operation because one of them coincides with the axis of the cylinder that has been formed. This coincidence of local and global geometries suggests that an \emph{orthotropic response in planes orthogonal to radial directions} be presumed for the material comprising a shell-like three-dimensional body intended to model zigzag and armchair CNTs. This presumption is central to the theory developed in \cite{Fa2}, that we briefly recapitulate here.

The elasticity tensor we pick is: 
\begin{equation}\label{constakl}
\begin{aligned}
\Co&=\Delta^{-1}\Big(E_1\Wb_1\otimes\Wb_1+
E_2\Wb_2\otimes\Wb_2+2G\,\Delta\, 
\Wb_3\otimes\Wb_3\\
&+E_1\nu_{21}(\Wb_1\otimes\Wb_2+\Wb_2\otimes\Wb_1)\Big),\quad \Delta:=1-\nu_{12}\,\nu_{21},
\end{aligned}
\end{equation}
%
where $E_1$ and $E_2$ are Young's moduli, $G$ a shear modulus, and
$\nu_{12}$, $\nu_{21}$ Poisson's coefficients; the relation that makes the choices of Young's and shear moduli interdependent is:
\begin{equation}\label{inter}
E_1\nu_{21}=E_2\nu_{12}=:\eta.
\end{equation}
The only nonnull components of $\Co$ are:
\[
\begin{aligned}
&\Co_{1111}=\Delta^{-1}E_1, \quad \Co_{2222}=\Delta^{-1}E_2,  \\
& \Co_{1122}=\Co_{2211}=\Delta^{-1}\eta,\quad \Co_{3333}=2G\,.
\end{aligned}
\]
\emph{We interpret $\Co$ as the elasticity tensor of a $(n,0)$-zigzag CNT}, and note that the dependence on the integer $n$ of all material moduli can be determined by means of a rather complex procedure detailed in \cite{BFPPG},  that needs not be summarized here. The elasticity tensor for a $(n,0)$-armchair CNT will be deduced from $\Co$ in the next subsection, where it becomes clear that switching the roles of the two achiral types of CNTs would not change anything substantial in our reasoning. 

\vskip 6pt
\noindent\textbf{Remark.} It follows from \eqref{bonstrain} and \eqref{constakl} that the stress field in $\Gc(\Sc, \eps)$ consists in a reactive part $\Sb^R$ and an active part $\Sb^A$:
\begin{equation}\label{reac}
\Sb=\Sb^R+\Sb^A,
\end{equation}
with
$$
\Sb^R= s_i^R\Vb_i \quad\textrm{and}\quad \Sb^A=\Co[\Eb(\ub)]=e_{i}(\ub)\,\Co[\Wb_i]\,.
$$
Note that
\[
\begin{aligned}
\Co[\Wb_1]&=\Delta^{-1}(E_1\Wb_1+\eta\Wb_2),\\
\Co[\Wb_2]&=\Delta^{-1}(E_2\Wb_2+\eta\Wb_1),\\
\Co[\Wb_3]&=2G\,\Wb_3\,,
\end{aligned}
\]
whence
\[
\begin{aligned}
\Sb^A=&
\Delta^{-1}\Big( \big(E_1e_1(\ub)+\eta e_2(\ub)\big)\Wb_1+\big(E_2 e_2(\ub)+\eta e_1(\ub)\big)\Wb_2+\\
&2G\,e_3(\ub)\,\Wb_3\Big).
\end{aligned}
\]
Note also that the \emph{strain energy} per unit volume $\Gc(\Sc, \eps)$ is:
\[
\begin{aligned}
w(\Eb):=&\frac{1}{2}\,\Eb\cdot\Co[\Eb]=\\
&\frac{1}{2}\,\big(E_1\,E_{11}^2+2\eta\,E_{11}E_{22}+ E_2\,E_{22}^2+2G\,E_{12}^2 \big).
\end{aligned}
\]
%

%

\subsection{The response of chiral CNSs}\label{C}
For chiral CNTs the local geometry of the material and the global geometry of  the  associated cylindrical shell cease to agree, in the sense that the orthotropy axes do not coincide anymore with the  chiral and roll-up axes. However, the \emph{anisotropic} response of a $(n,m)$-chiral CNT can be induced, alternatively, from the orthotropic response of the corresponding $(n,0)$-zigzag or $(n,n)$-armchair CNT.

Let the \emph{tensor product} $\boxtimes$ of any two second-order tensors $\Ab,\Bb$ be the fourth-order tensor defined as follows by its linear action on the collection of second-order tensors:
\[
\Ab\boxtimes\Bb[\Cb]:=\Ab\Cb\Bb^T,\quad\textrm{for each second-order tensor}\; \Cb.
\]
For $\Qb$ an orthogonal tensor, the tensor product $\Qb\boxtimes\Qb=:\Qo$ defines the fourth-order tensor that delivers the \emph{orthogonal conjugate} with respect to $\Qb$ of a given second-order tensor $\Cb$: 
\[
\Qo\Cb=\Qb\Cb\Qb^T.
\]

Now, let $\Qb$ be a rotation of $\psi$ radians about an axis parallel to $\eb_3$:
\begin{equation}\label{rotaz}
\begin{aligned}
\Qb=\widehat\Qb(\psi,\eb_3):=&\cos\psi\,(\eb_1\otimes\eb_1+\eb_2\otimes\eb_2)+\\
&-\sin\psi\,(\eb_1\otimes\eb_2-\eb_2\otimes\eb_1)+\eb_3\otimes\eb_3.
\end{aligned}
\end{equation}
For such a rotation, consider the fourth-order tensor
\begin{equation}\label{rotko}
\widetilde\Co:=\Qo^T\Co\Qo.
\end{equation}
Note that the Cartesian components of $\widetilde\Co$ with respect to the orthonormal frame $(\eb_1,\eb_2,\eb_3)$, namely,
\begin{equation}\label{1111}
\widetilde{\Co}_{ijhk}=Q_{li}Q_{mj}Q_{nh}Q_{pk}\Co_{lmnp},
\end{equation}
 are identical to the Cartesian components with the same indices of $\Co$ with respect to the orthonormal frame 
\[
(\widetilde\eb_1,\widetilde\eb_2,\eb_3),\quad \widetilde\eb_{\alpha}=\Qb\eb_\alpha,\;\,(\alpha=1,2);
\]
for example, 
\[
\begin{aligned}
\widetilde{\Co}_{1111}&=Q_{11}^4\Co_{1111}+Q_{21}^{4}\Co_{2222}+2Q_{11}^2Q_{21}^2(\Co_{1122}+\Co_{1212})\\
&=\Co_{1111}\cos^4\psi+\Co_{2222}\sin^4\psi+\\
&\hspace{0.5cm}2(\Co_{1122}+\Co_{1212})\sin^2\psi\cos^2\psi.
\end{aligned}
\]

Chirality enters \eqref{rotko} in two ways: because $\Co$ is the elasticity tensor of a $(n,0)$-zigzag CNT, whose representation is given in \eqref{constakl}; and because we compose the mapping $\widehat\Qb(\cdot,\eb_3)$ introduced in \eqref{rotaz} with the function
\begin{equation}
\psi=\widehat\psi(n,m):=\left\{\begin{array}{c}
0, \quad\qquad\;\; {\rm if}\; m=0\\
\frac{\pi}{3}+\varphi(n,m) \quad {\rm if}\;m\in(0,n] ,\end{array}\right.
\end{equation}

that is to say, in view of \eqref{chirang},
\begin{equation}
\widehat\psi(n,m)=\left\{\begin{array}{c}
0, \qquad\qquad\quad\,\quad {\rm if}\; m=0\\
\arctan\left(\sqrt{3}\,\frac{n+m}{n-m}  \right) \quad {\rm if}\;
m\in(0,n] .\end{array}\right.
\end{equation}
On denoting by
\[
\widetilde\Qb(n,m):=\widehat\Qb(\widehat\psi(n,m),\eb_3)
\]
the rotation mapping associated to a given $(n,m)$-chiral CNT by way of the composition operation we just mentioned, we get:
\begin{itemize}
\item for $m=0$ ($\psi=\varphi=0$, $\widetilde\Qb(n,0)=\Ib$, the identity tensor), $\widetilde\Co=\Co$; 
\item for $n=m$ ($\psi=\pi/2,\;\varphi=\pi/6$, $\widetilde\Qb(n,n)=\eb_2\otimes\eb_1-\eb_1\otimes\eb_2+\eb_3\otimes\eb_3$), $\widetilde\Co$ becomes the elasticity tensor of a $(n,n)$=armchair CNT, for which, according to \eqref{1111}, $$\widetilde\Co_{1111}=\Co_{2222}, \quad\widetilde\Co_{2222}=\Co_{1111},\;\, \textrm{etc.};$$
\item for $m\in (0,n)$ ($\psi\in(0,\pi/2),\;\,\varphi\in (0,\pi/6)$), $\widetilde\Co$ captures the elastic response for intermediate
chiralities.
\end{itemize}
\vskip 4pt
\noindent\textbf{Remarks.}  1. The following figure is meant to help visualizing the action of a rotation by $\psi$ about the axis $\eb_3$.
\begin{figure}[h]
\centering
\includegraphics[scale=0.5]{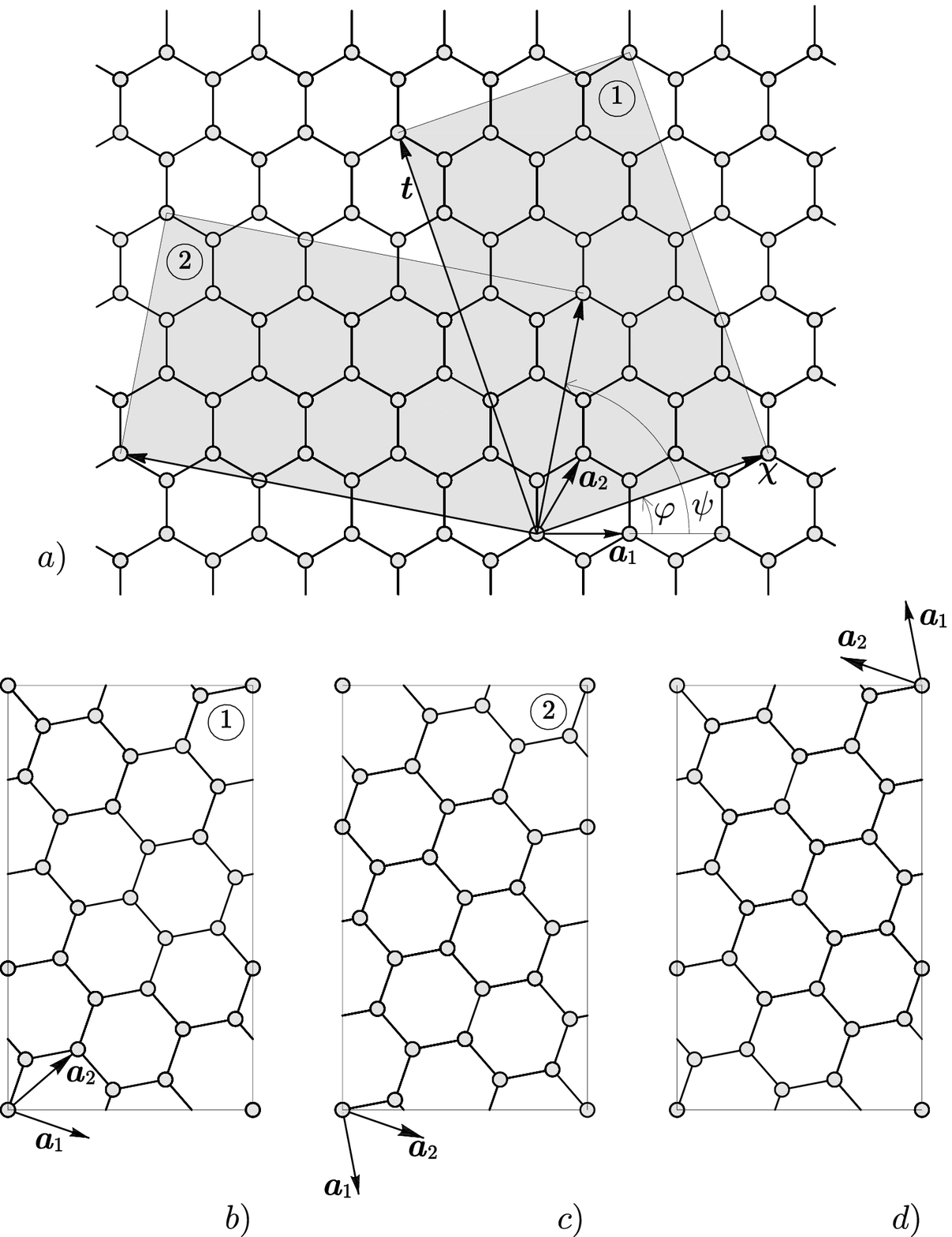}
\caption{}
\label{simmetrie}
\end{figure}
In Fig. \ref{simmetrie}$a)$, the graphene portion corresponding to a $(2,1)$-CNT, whose chiral angle is $\varphi$, is marked ${\bigcirc \kern-7pt{\scriptstyle 1}}\;$; its rotation by an angle $\pi/3$ gives portion ${\bigcirc \kern-7pt{\scriptstyle 2}}\;$; the angle $\psi$ is shown. In Fig.s \ref{simmetrie}$b)$ and \ref{simmetrie}$c)$, ${\bigcirc \kern-7pt{\scriptstyle 1}}\;$ and  ${\bigcirc \kern-7pt{\scriptstyle 2}}\;$ have been rotated so as to have parallel roll-up axes;  in Fig. \ref{simmetrie}$d)$,  ${\bigcirc \kern-7pt{\scriptstyle 2}}\;$ has been  further rotated by $\pi$ radians, so as to make evident the fact that its atomic arrangement is the same as that of ${\bigcirc \kern-7pt{\scriptstyle 1}}\;$. But, such a final rotation belongs to the symmetry group of an orthotropic material. Thus,  the CNTs obtained by rolling up the graphene portions ${\bigcirc \kern-7pt{\scriptstyle 1}}\;$ and  ${\bigcirc \kern-7pt{\scriptstyle 2}}\;$ depicted in Fig. \ref{simmetrie}$a)$ have the same mechanical response.
\vskip 6pt

\noindent 2. Let
$\Sb=\Co[\Eb]$, $\widetilde\Eb=\Qo^T[\Eb]$ and $\widetilde\Sb=\widetilde\Co[\widetilde\Eb]$. Then, it follows from \eqref{rotko} that
\begin{equation}\label{rotko1}
\begin{aligned}
\widetilde\Sb=&\big(\widetilde\Co[\widetilde\Eb]=(\Qo^T\Co\Qo)[\widetilde\Eb]=(\Qo^T\Co)[\Qo[\widetilde\Eb]]=\\
&\Qo^T[\Co[\Qo[\Qo^T[\Eb]]]=\Qo^T[\Co[\Eb]]=\big)\Qo^T[\Sb].
\end{aligned}
\end{equation}
Now, let $\tb:=\Sb\nb$ be the active traction vector relative to a plane of normal $\nb$ through a typical point of the shell associated to a given $(n,0)$-zigzag CNT, whose elastic response is described by $\Co$, when $\Eb$ measures the strain at that point.  Then, for $\widetilde\nb=\Qb^T\nb$ and $\widetilde\tb:=\widetilde\Sb\widetilde\nb$,  \eqref{rotko1} implies that
\[
\widetilde\tb=\Qb^T\tb,
\]
that is to say, that the a rotation by $\widetilde\Qb(n,m)$ of $\tb$ gives the active traction vector at a plane of normal $\widetilde\nb$ through a typical point of the shell associated to a $(n,m)$-chiral CNT, whose elastic response is described by $\widetilde\Co$, when the strain at that point is $\widetilde\Eb$.
\section{Thickness and radius}\label{RT}
As anticipated in the Introduction, an analysis detailed in  \cite{BFPPG} leads  to precise definitions for the \emph{effective thickness} and \emph{effective radius} of the CNS to be associated to either a $(n,n)$-armchair CNT or $(n,0)$-zigzag CNT. The figures here below are taken from \cite{BFPPG}.  
 
\begin{figure}[h]
\centering
\includegraphics[scale=0.7]{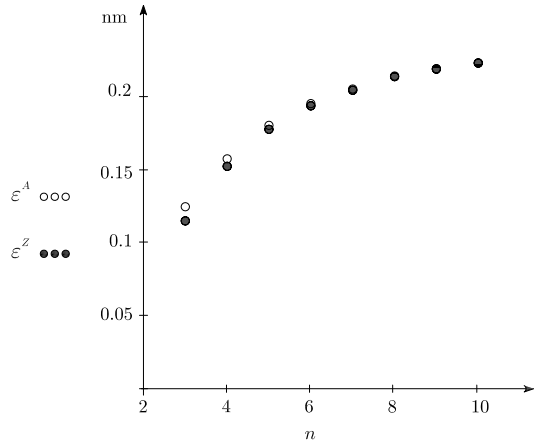}
\caption{The effective thickness of A-$(\circ)$ and Z-$(\bullet)$ CNSs.} \label{epsilon}
\end{figure}

As to effective thickness, we see in Fig. \ref{epsilon} 
that it depends on the parameter $n$ pretty much in the same way for both armchair- and zigzag-CNTs; and that the small differences existing for $n$ small disappear very quickly. Therefore,  for the CNS associated to a given $(n,m)$-chiral CNT we take the effective thickness of a $(n,0)$-CNT.

As to the effective radius, 
\begin{figure}[h]
\centering
\includegraphics[scale=0.7]{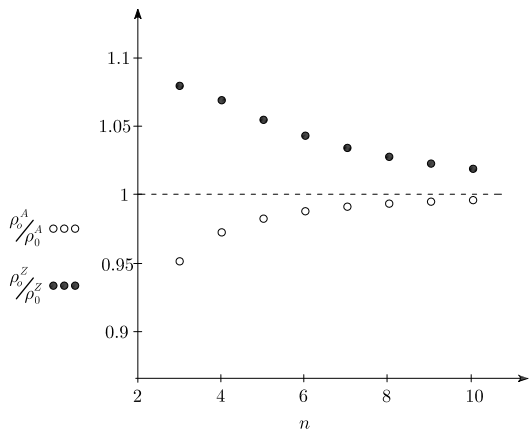}
\caption{Ratio of effective-to-nominal radii of A-$(\circ)$ and Z-$(\bullet)$ CNSs.}
\label{raggio}
\end{figure}
Fig. \ref{raggio}
shows that it tends to equal the nominal radius when $n$ grows big. 
Therefore, for a $(n, m)$-chiral CNT we take the effective radius to be equal to its 
nominal radius, given by \eqref{geonec}.

\section{Balance equations}\label{baleq}
\begin{figure}[h]
\centering
\includegraphics[scale=0.7]{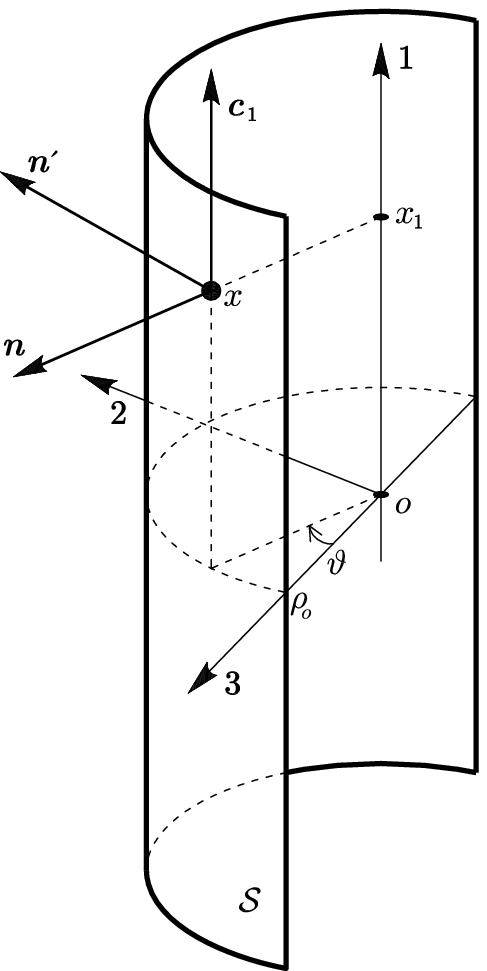}
\caption{A portion of the model surface of a right cylindrical
shell.} \label{cyl}
\end{figure}

Consider the shell-shaped region $\Gc(\Sc, \eps)$, of constant thickness $(2\varepsilon)$, modeled over the cylindrical surface $\Sc$. For $(x_1, \vartheta)$ the cylindrical coordinates of a point $x\in\Sc$ with respect to an origin $o$, a point $p\in\Gc(\Sc,\eps)$ has position vector
$$
\pb:=p-o=x-o+\zeta\nb(x),\quad x\in\Sc,\;\,\zeta\in
I:=(-\varepsilon,+\varepsilon);
$$
as shown in Fig. \ref{cyl},
 $\nb(x)$ is the outer unit normal to $\Sc$ at $x$, and  $|\zeta|$ is the distance of $p$
from $x$. The admissible displacements of $\Gc(\Sc, \eps)$, those that satisfy the Kirchhoff--Love constraint  \eqref{KL} with $\eb_3\equiv\nb$, have the form:
\begin{equation}\label{discil}
\begin{aligned}
\ub(x_1,&\vartheta, \zeta)=\\
&\big(a_{1}-\zeta w,_1 \big)\cb_1+\left(a_{2}-\frac{\zeta}{\rho_o}(a_{2}-w,_2) \right)\nb'+w\nb,
\end{aligned}
\end{equation}
parameterized by two fields defined over $\Sc$, the vector field $\ab$  everywhere orthogonal to $\Sc$, and the scalar field $w$
(here and henceforth, $(\cdot),_1$ and $(\cdot),_2$ denote differentiation with respect to $x_1$ and $\vartheta$, respectively;
when a field depends only on the latter variable, we prefer to write $(\cdot)^\prime$ instead of $(\cdot),_2$).

\subsection{Principles of Virtual Power, three- and two-dimensional}
Just as we did in \cite{Fa2}, we state the equilibrium of $\Gc(\Sc, \eps)$ by formulating a three-dimensional Principle of
Virtual Powers having two special traits:
\begin{itemize}
\item [(i)] \emph{body parts} are tubular neighborhoods of thickness $2\varepsilon$ of open subsets of $\Sc$;
\item [(ii)] \emph{virtual velocities} are consistent with the representation \eqref{discil} of admissible displacements, and hence have the form:
\[
\vb=\overset{(0)}\vb+\zeta\overset{(1)}\vb,
\]
with
\begin{equation}\label{vars}
\begin{aligned}
&\overset{(0)}\vb=v_1\cb_1+v_2\nb'+v_3\nb,\\
&\overset{(1)}\vb=-v_3,_1\cb_1+\rho_o^{-1}(v_2-v_3,_2)\nb',
\end{aligned}
\end{equation}
the scalar fields $v_i=v_i(x_1,\vartheta)$ $(i=1,2,3)$ being
compactly supported in $\Sc$.
\end{itemize}
Under these assumptions, we postulate that
\begin{equation}\label{PVP}
\int_P\Sb\cdot\nabla\vb=\int_P\db_o\cdot\vb+\int_{\partial P}\cb_o\cdot\vb\,,
\end{equation}
for all parts $P\equiv\Pc\times(-\varepsilon,+\varepsilon)$ of $\Gc(\Sc, \eps)$, where, in accord with (i), $\Pc$ is a part of $\Sc$, and for all virtual velocity fields specified under (ii). Here $\Sb$ denotes the restriction to $P$ of the \emph{active} stress field $\Sb^A$ in $\Gc(\Sc, \eps)$,\footnote{It follows from assumption (ii) that $\Sb^R\cdot\nabla\vb\equiv 0$ in $\Gc(\Sc, \eps)$.}
 and  $(\db_o,\cb_o)$ denote, respectively, the \emph{distance
force} for unit volume and the \emph{contact force} per unit area
exerted on $P$ by its own complement with respect to $\Gc(\Sc, \eps)$
and/or by the environment of the latter.

It is the matter of straightforward calculations to deduce from \eqref{PVP} the following two-dimensional Principle of Virtual Power:
\begin{equation}\label{2Dvp}
\begin{aligned}
&\int_\Pc\big(
\Fb\cdot\nabla\overset{(0)}\vb+\Mb\cdot\nabla\overset{(1)}\vb
+\fb^{(3)}\cdot\overset{(1)}\vb\big)=\\
&\int_\Pc(\qb_o\cdot\overset{(0)}\vb+
\rb_o\cdot\overset{(1)}\vb)+\int_{\partial\Pc}(\lb_o\cdot\overset{(0)}\vb+\mb_o\cdot\overset{(1)}\vb),
\end{aligned}
\end{equation}
for all parts $\Pc$ of $\Sc$ and for all virtual fields $\overset{(0)}\vb,\overset{(1)}\vb$ as in \eqref{vars}. 
The stress-like fields are here the \emph{force tensor} $\Fb$,
the \emph{moment tensor} $\Mb$, and the \emph{shear vector}
$\fb^{(3)}$:  
\begin{equation}\label{2Ds}
\begin{aligned}
\Fb(x)&:=\Big(\int_I\alpha(x,\zeta)\Sb(x,\zeta)\gb^\alpha(x,\zeta) d\zeta\Big)\otimes\eb_\alpha,\\
\Mb(x)&:=\Big(\int_I\alpha(x,\zeta)\zeta\Sb(x,\zeta)\gb^\alpha(x,\zeta)
d\zeta\Big)\otimes\eb_\alpha,
\end{aligned}
\end{equation}
and
\begin{equation}\label{2Dsh}
\fb^{(3)}(x):=\int_I\alpha(x,\zeta)\Sb(x,\zeta)\nb(x) d\zeta.\footnote{For the details of the integration method, the reader is referred to \cite{Fa2}. Here it is sufficient to declare that $\alpha:=\det\Ab$, with $\Ab$ and $\Bb$ the shifter tensors correlating the covariant and contravariant bases and vice versa.}
\end{equation}
The force-like fields $\qb_o$, $\rb_o$ are, respectively, the two-dimensional \emph{distance force}
and \emph{distance couple} per unit area, and $\lb_o$, $\mb_o$ are the
\emph{contact force} and \emph{contact couple} per unit length:
\begin{equation}\label{forceload}
\begin{aligned}
\qb_o(x)&:=\int_I\alpha(x,\zeta)\db_o(x,\zeta)\,d\zeta+\\
&\hspace{0.5cm}\alpha^+(x)\cb_o^+(x)+\alpha^-(x)\cb_o^-(x),\\
\rb_o(x)&:=\int_I\alpha(x,\zeta)\zeta\db_o(x,\zeta)\,d\zeta+\\
&\hspace{0.5cm}\varepsilon\big(\alpha^+(x)\cb_o^+(x)-\alpha^-(x)\cb_o^-(x)\big),
\end{aligned}
\end{equation}
$$
\lb_o(x):=\int_I\cb_o(x,\zeta)d\,\zeta,\quad
\mb_o(x):=\int_I\zeta\cb_o(x,\zeta)\,d\zeta.
$$
In the above-mentioned formulas we put
$$
\alpha=1+\frac{\zeta}{\rho_o}\,,\quad
\alpha^\pm:=\alpha(x,\pm\varepsilon), \quad
\cb_o^\pm:=\cb_o(x,\pm\varepsilon), 
$$
and
\begin{equation}
\begin{aligned}
&\eb_1=\gb^1=\cb_1,\\
&\eb_2=\rho_o\nb',\\
&\eb_3=\gb^3=\nb,\\
&\gb^2=\left(1+\frac{\zeta}{\rho_o}\right)^{-1}\rho_o^{-1}\nb'.\\
\end{aligned}
\end{equation}

\subsection{Field equations and boundary conditions}
The field equations and the
boundary conditions of our theory of CNSs are deduced from the two-dimensional Principle of Virtual Power \eqref{2Dvp} by exploiting the quantifications inherent to its formulation.

We find that the following field equations must be satisfied at each point of $\Sc$:
\begin{equation}\label{equicil}
\begin{aligned}
&F_{11,1}+\rho_o^{-1}F_{12,2}+q_{o1}=0,\\
&(F_{21}+\rho_o^{-1}M_{21,1})+\rho_o^{-1}(F_{22}+\rho_o^{-1}M_{22,2})+q_{o2}+\rho_o^{-1}r_{o2}=0,\\
&M_{11},_{11}+\rho_o^{-1}(M_{12}+M_{21})_{,12}+\frac{1}{\rho_o^2}M_{22},_{22}+\\
&\hspace{2cm}-\rho_o^{-1}F_{22}+q_{o3}+r_{o1},_{1}+\rho_o^{-1}r_{o2},_{2}=0,\\
\end{aligned}
\end{equation}
where $F_{\alpha\beta}$ and $M_{\alpha\beta}$ are the physical components of
the force and moment tensors and $q_{o\alpha}$ and $r_{o\alpha}$ are
the physical components of the distance force and couple. 

We also find that, at a boundary point belonging to a directrix of the cylinder\footnote{Other types of boundary curves are considered in \cite{Fa2}.}, the admissible boundary conditions
must consist of a list of mutually exclusive assignments of the
one or the other element of the following five power-conjugate
pairs:
\begin{equation}\label{conto1}
\begin{aligned}
&(F_{11},  a_{1}),\;
(F_{21}+\rho_o^{-1}M_{21},a_{2}),\\
&(F_{31}-\rho_o^{-1}M_{21},_{2}, w),\;(M_{11}, w,_1).
\end{aligned}
\end{equation}

\subsection{Axisymmetric problems}
A boundary-value problem for a cylindrical shell is
\emph{axisymmetric} if the load and confinement data induce
equilibrium displacement fields whose components 
are all independent of the circumferential coordinate $\vartheta$,
that is to say, if
\begin{equation}\label{cilcil}
u_{1}=a_{1}-\zeta w', \quad u_{2}=\left(1+\frac{\zeta}{\rho_o}
\right)a_{2}, \quad u_{3}=w,
\end{equation}
where a prime denotes differentiation with respect to $x_1$, the
only space variable from which all of the parameter fields
$a_{1},a_{2}$, and $w$, may depend. When the displacement field
has the form \eqref{cilcil}, the strain components take the
form:
\begin{equation}\label{strainax}
\begin{aligned}
E_{11}&=a_{1}'-\zeta w''\,,\\
E_{12}&=E_{21}=\frac{1}{2}\left( 1+\frac{\zeta}{\rho_o}
\right)a_{2}',\\
E_{22}&=\left(\rho_o\left(
1+\frac{\zeta}{\rho_o} \right) \right)^{-1}w.\\
\end{aligned}
\end{equation}

When modeling a chiral CNT as a linearly elastic CNS, the elasticity tensor is given by \eqref{rotko} and hence the active stress induced by a strain of type \eqref{strainax} in $\Gc(\Sc, \eps)$ is:
\begin{equation}\label{S11}
\begin{aligned}
S_{ij}=a_{ij}E_{11}+b_{ij}E_{22}+c_{ij}E_{12},
\end{aligned}
\end{equation}
where
\begin{equation}
a_{ij}:=\widetilde\Co_{ij11}, \quad b_{ij}:=\widetilde\Co_{ij22},\quad c_{ij}:=2\,\widetilde\Co_{ij11},
\end{equation}
and
\begin{align}
&a_{11}:=\frac{E_1}{\Delta}\cos^4\psi+\frac{E_2}{\Delta}\sin^4\psi+\frac{2\eta}{\Delta} \sin^2\psi\cos^2\psi+\\
&\hspace{1cm}G\sin^22\psi,\\
&b_{11}:=\frac{\eta}{\Delta}\cos^4\psi+\frac{\eta}{\Delta}\sin^4\psi+\left(\frac{E_1}{\Delta}+\frac{E_2}{\Delta}\right)\sin^2\psi\cos^2\psi\\
&\hspace{1cm}-G\sin^22\psi,\\
&c_{11}:=2G\sin2\psi\cos2\psi-\left(\frac{E_1}{\Delta}(1-\nu_{21})\cos^2\psi+\right.\\
&\hspace{1cm}\left.-\frac{E_2}{\Delta}(1-\nu_{12})\sin^2\psi\right)\sin2\psi,\\
&a_{22}:=\frac{\eta}{\Delta}\sin^4\psi+\frac{\eta}{\Delta}\cos^4\psi+\left(\frac{E_1}{\Delta}+\frac{E_2}{\Delta}\right)\sin^2\psi\cos^2\psi\\
&\hspace{1cm}-G\sin^22\psi,\\
&b_{22}:=\frac{E_1}{\Delta}\sin^4\psi+\frac{E_2}{\Delta}\cos^4\psi+\frac{2\eta}{\Delta}\sin^2\psi\cos^2\psi+\\
&\hspace{1cm}G\sin^22\psi,\\
&c_{22}:=-\left(2G\sin2\psi\cos2\psi+\left(\frac{E_1}{\Delta}(1-\nu_{21})\cos^2\psi+\right.\right.\\
&\hspace{1cm}\left.\left.-\frac{E_2}{\Delta}(1-\nu_{12})\sin^2\psi\right)\sin2\psi\right),\\
&a_{12}:=\frac{1}{2}G\sin4\psi-\left(\frac{E_1}{\Delta}-\frac{\eta}{\Delta}  \right)\cos^3\psi\sin\psi+\\
&\hspace{1cm}-\left(\frac{\eta}{\Delta}-\frac{E_2}{\Delta} \right)\sin^3\psi\sin\psi,\\
&b_{12}:=-\frac{1}{2}G\sin4\psi-\left(\frac{\eta}{\Delta}-\frac{E_2}{\Delta} \right)\cos^3\psi\sin\psi+\\
&\hspace{1cm}-\left(\frac{E_1}{\Delta}-\frac{\eta}{\Delta}  \right)\sin^3\psi\sin\psi,\\
&c_{12}:=2G\cos^22\psi+\frac{1}{2}\left(\frac{E_1}{\Delta}(1-\nu_{21})+\right.\\
&\hspace{1cm}\left.\frac{E_2}{\Delta}(1-\nu_{12})\right)\sin^22\psi.
\end{align}

\remark
In the zigzag case, $\psi=0$ and \eqref{S11} yields:
\begin{equation}
\begin{aligned}
&S_{11}=\frac{E_1}{\Delta}\big(E_{11}+\nu_{21}E_{22}\big),\\
&S_{22}=\frac{E_2}{\Delta}\big(E_{22}+\nu_{12}E_{11}\big),\\
&S_{12}=2GE_{12};
\end{aligned}
\end{equation}
in the armchair case, when $\psi=\pi/2$, we have:
\begin{equation}
\begin{aligned}
&S_{11}=\frac{E_2}{\Delta}\big(E_{11}+\nu_{12}E_{22}\big),\\
&S_{22}=\frac{E_1}{\Delta}\big(E_{22}+\nu_{21}E_{11}\big),\\
&S_{12}=2GE_{12}.
\end{aligned}
\end{equation}

\vskip 8pt

When a problem is axisymmetric, the shear vector defined in \eqref{2Dsh} is everywhere null, whereas the force and moment tensors defined in \eqref{2Ds} have the following expressions in terms of the parameter fields $\,a_{1},a_{2}$ and $w$, from which the displacement field \eqref{cilcil} depends:
\begin{subequations}
\begin{align}
F_{11}&=-\frac{2}{3}\frac{\varepsilon^3}{\rho_o}a_{11}w''+2\frac{\varepsilon}{\rho_o}b_{11}w+2\varepsilon a_{11}a_{1}'+\\
&\varepsilon\left(1+\frac{1}{3}\frac{\varepsilon^2}{\rho_o^2}\right)c_{11}a_{2}',\label{F11}\\
F_{22}&=2\varepsilon a_{22}a_{1}'+2\frac{\varepsilon}{\rho_o}\frac{1}{2\frac{\varepsilon}{\rho_o}}\log\frac{1+\frac{\varepsilon}{\rho_o}}{1-\frac{\varepsilon}{\rho_o}} b_{22}w+\\
&2\varepsilon c_{22}a_{2}',\label{F22}\\
F_{12}&=2\varepsilon a_{12}a_{1}'+2\frac{\varepsilon}{\rho_o}\frac{1}{2\frac{\varepsilon}{\rho_o}}\log\frac{1+\frac{\varepsilon}{\rho_o}}{1-\frac{\varepsilon}{\rho_o}}b_{12}w+2\varepsilon c_{12}a_{2}',\label{F12}\\
F_{21}&=2\varepsilon a_{12}\left(a_{1}'-\frac{1}{3}\frac{\varepsilon^2}{\rho_o}w''\right)+2\frac{\varepsilon}{\rho_o} b_{12} w+\\
&\varepsilon\left(1+\frac{1}{3}\frac{\varepsilon^2}{\rho_o^2}\right)c_{12}a_{2}',\label{F21}\\
M_{11}&=-\frac{2}{3}\frac{\varepsilon^3}{\rho_o}\Big( a_{11}(\rho_ow''-a_{1}')-c_{11}a_{2}') \Big),\label{M11}\\
M_{12}&=-\frac{2}{3}\varepsilon^3a_{12}w''+2\varepsilon\left( 1-\frac{1}{2\frac{\varepsilon}{\rho_o}}\log\frac{1+\frac{\varepsilon}{\rho_o}}{1-\frac{\varepsilon}{\rho_o}}\right)b_{12}w+\\
&\frac{1}{3}\frac{\varepsilon^3}{\rho_o}c_{12}a_{2}',\label{M12}\\
M_{21}&=-\frac{2}{3}\frac{\varepsilon^3}{\rho_o}\Big( a_{12}(\rho_ow''-a_{1}')-c_{12}a_{2}') \Big),\label{M21}\\
M_{22}&=-\frac{2}{3}\varepsilon^3a_{22}w''+2\varepsilon\left( 1-\frac{1}{2\frac{\varepsilon}{\rho_o}}\log\frac{1+\frac{\varepsilon}{\rho_o}}{1-\frac{\varepsilon}{\rho_o}}\right)b_{22}w+\\
&\frac{1}{3}\frac{\varepsilon^3}{\rho_o}c_{22}a_{2}'.\label{M22}
\end{align}
\end{subequations}

\section{The torsion problem}
Let us consider a chiral CNS subject to a distribution of end
tractions statically equivalent to two mutually balancing torques
of magnitude
\begin{equation}\label{tigrande}
T=(2\pi\rho_o^2)t, \quad\textrm{with}\;\, t=O(\varepsilon).
\end{equation} 
The general field equations \eqref{equicil} reduce to:
\begin{equation}\label{equil}
\begin{aligned}
&F_{11}'=0,\\
&(F_{21}+\rho_o^{-1}M_{21})'=0,\\
&M_{11}''-\rho_{o}^{-1}F_{22}=0,\\
\end{aligned}
\end{equation}
holding in the interval $(-l,+l)$; the boundary conditions prevailing at 
$\pm l$ are (cf. \eqref{conto1}):
\begin{equation}\label{bc}
\begin{aligned}
&F_{11}=0, \quad M_{11}=0, \quad M_{11}'=0, \quad
F_{21}+\rho_o^{-1}M_{21}=t\,.
\end{aligned}
\end{equation}

Equations $\eqref{equil}_2$ and $\eqref{bc}_4$ allow to conclude that $$F_{21}+\rho_o^{-1}M_{21}=t\quad \textrm{in}\;[-l,l],$$ a condition which, with the use of \eqref{F21} and \eqref{M21}, can be written as follows:

\begin{equation}
\begin{aligned}
&-\frac{4}{3}\frac{\varepsilon^3}{\rho_o}a_{12}w''+2\frac{\varepsilon}{\rho_o}b_{12}w+2\varepsilon\left(1+\frac{1}{3}\frac{\varepsilon^2}{\rho_o^2}\right)a_{12}a_{1}'+\\
&\varepsilon\left(1+\frac{\varepsilon^2}{\rho_o^2} \right)c_{12}a_{2}'=t
\end{aligned}
\end{equation}
This last relation imply an expression for $a_{2}'$ that will be useful later, namely,
\begin{equation}\label{Ga2}
\begin{aligned}
a_{2}'=&\left(\varepsilon\left(1+\frac{\varepsilon^2}{\rho_o^2} \right)c_{12}  \right)^{-1}\left(t+\frac{4}{3}\frac{\varepsilon^3}{\rho_o}a_{12}w''- 2\frac{\varepsilon}{\rho_o}b_{12}w+\right.\\
&\left.-2\varepsilon\left(1+\frac{1}{3}\frac{\varepsilon^2}{\rho_o^2}\right)a_{12}a_{1}' \right).
\end{aligned}
\end{equation}

Equation $\eqref{equil}_1$, together with the boundary condition $\eqref{bc}_1$, allows to conclude that $F_{11}=0$ in $[-l,l]$, a condition that, in the light of \eqref{F11}, reads:
\begin{equation}\label{F11e}
\begin{aligned}
-\frac{2}{3}\frac{\varepsilon^3}{\rho_o}a_{11}w''+2\frac{\varepsilon}{\rho_o}b_{11}w+&2\varepsilon a_{11}a_{1}'+\\
&\varepsilon\left(1+\frac{1}{3}\frac{\varepsilon^2}{\rho_o^2}\right)c_{11}a_{2}'=0.
\end{aligned}
\end{equation}
On substituting the expression \eqref{Ga2} for $a_{2}'$, \eqref{F11e} yields an expression for $a_{1}'$ in terms of the function $w$, its derivatives, and the datum $t$:
\begin{equation}\label{a1}
\begin{aligned}
a_{1}'=&\left(a_{11}- \frac{\left(1+\frac{1}{3}\frac{\varepsilon^2}{\rho_o^2}\right)^2}{\left(1+\frac{\varepsilon^2}{\rho_o^2}\right)}a_{12} \right)^{-1}\left(\left(\frac{1}{3}a_{11}-\frac{2}{3}\frac{1+\frac{1}{3}\frac{\varepsilon^2}{\rho_o^2}}{1+\frac{\varepsilon^2}{\rho_o^2}}\frac{c_{11}}{c_{12}}a_{12}\right)\times\right.\\
&\left.\times\frac{\varepsilon^2}{\rho_o}w''+ +\left(\frac{1+\frac{1}{3}\frac{\varepsilon^2}{\rho_o^2}}{1+\frac{\varepsilon^2}{\rho_o^2}}\frac{c_{11}}{c_{12}}b_{12}-b_{11} \right)\frac{w}{\rho_o}+\right.\\ &\left.-\frac{t}{2\varepsilon}\frac{1+\frac{1}{3}\frac{\varepsilon^2}{\rho_o^2}}{1+\frac{\varepsilon^2}{\rho_o^2}}\frac{c_{11}}{c_{12}}\right).
\end{aligned}
\end{equation}
One last equilibrium equation remains, that is, $\eqref{equil}_3$. Recalling \eqref{F22}, \eqref{M11}, \eqref{Ga2}, and \eqref{a1}, that equation yields an ODE for the only unknown $w$:
\begin{equation}\label{eqdiff}
\begin{aligned}
c_1w''''+c_2w''+c_3w+c_4 t=0,
\end{aligned}
\end{equation}
whose coefficients have lengthy expressions, that we relegate in the final Appendix, in terms of $a_{\alpha\beta}, b_{\alpha\beta},c_{\alpha\beta}$ $(\alpha,\beta=1,2)$, and the geometric parameters $\rho_o,\varepsilon$. 

Given the problem's symmetries, we look for an even solution of \eqref{eqdiff}. It is not difficult to see that the most general even solution of the homogeneous equation associated to \eqref{eqdiff} has the form:
\begin{equation}\label{wh}
\begin{aligned}
w_{h}(x_1)=t\big(&k_1\big(\exp(\alpha_1x_1)+\exp(-\alpha_1x_1) \big)+\\
&k_2\big(\exp(\alpha_2x_1)+\exp(-\alpha_2x_1) \big)\big),
\end{aligned}
\end{equation}
where
\begin{equation}
\begin{aligned}
&\alpha_1^2:=-\frac{1}{c_1}\left(c_2+\sqrt{c_2^2-4c_1c_3}  \right), \\ &\alpha_2^2:=-\frac{1}{c_1}\left(c_2-\sqrt{c_2^2-4c_1c_3}  \right).
\end{aligned}
\end{equation}
With this, we write:
\begin{equation}
w(x_1)=w_h(x_1)+w_p, \quad w_p:=-\frac{c_4}{c_3}\,t,
\end{equation}
with $w_p$ the constant solution of \eqref{eqdiff}.

With a view to determining the coefficients $k_1$ and $k_2$, we firstly return to the boundary conditions $\eqref{bc}_{1,4}$, that, when combined with \eqref{Ga2} and \eqref{a1}, furnish:
\begin{equation}\label{bound}
\begin{aligned}
a_{1}'(l)=A_1w''(l)+B_1w(l)+C_1t,\\
a_{2}'(l)=A_2w''(l)+B_2w(l)+C_2t
\end{aligned}
\end{equation}
(the lengthy expressions of coefficients $A_\alpha,B_\alpha,C_\alpha$ $(\alpha=1,2)$ are found in the Appendix).
Secondly,  by differentiating $\eqref{Ga2}$ and $\eqref{a1}$ and invoking continuity up to the boundary of the resultant expression, we obtain that
\begin{equation}\label{bound1}
\begin{aligned}
a_{1}''(l)=A_1w'''(l)+B_1w'(l),\\
a_{2}''(l)=A_2w'''(l)+B_2w'(l).
\end{aligned}
\end{equation}
Thirdly, we note that, with an use of \eqref{M11}, conditions $\eqref{bc}_{2,3}$ can be given the form:
\begin{equation}\label{sysconst}
\begin{aligned}
&a_{11}\big(\rho_ow''(l)-a_{1}'(l)\big)-c_{11}a_{2}'(l)=0,\\
&a_{11}\big(\rho_ow'''(l)-a_{1}''(l)\big)-c_{11}a_{2}''(l)=0,
\end{aligned}
\end{equation}
a system of equations that, on taking \eqref{bound} and \eqref{bound1} into account, determines the constants $k_\alpha$ in \eqref{wh}:
\begin{equation}
\begin{aligned}
&k_1=\widetilde{k}_1^{-1}\left(\alpha _2  \exp(\alpha _1 l) \left(\exp(2 \alpha _2 l)-1\right) \left(a_{11} C_1+c_{11}
   C_2\right)\right),\\
&k_2=\widetilde{k}_1^{-1}\left(\alpha _1  \exp(\alpha _2 l) \left(\exp(2 \alpha _1 l)-1\right) \left(a_{11} C_1+c_{11}
   C_2\right)\right),
\end{aligned}
\end{equation}
with
\begin{equation}
\begin{aligned}
&\widetilde{k}_1:=\left(\alpha _1 \left(\exp(2 \alpha _1 l)-1\right) \left(\exp(2 \alpha _2
   l)+1\right)+\right.\\
   &\hspace{1cm}\left.-\alpha _2 \left(\exp(2 \alpha _1 l)+1\right) \left(\exp(2 \alpha _2 l)-1\right)\right)\times\\
   &\hspace{1cm} \left(a_{11} \left(\alpha _1^2 \left(A_1-\rho_o \right)+B_1\right)+c_{11} \left(\alpha _1^2
   A_2+B_2\right)\right),\\
&\widetilde{k}_2:=\left(\alpha _2 \left(\exp(2 \alpha _1 l)+1\right) \left(\exp(2 \alpha _2 l)-1\right)-\right.\\
&\hspace{1cm}\left.\alpha _1 \left(\exp(2 \alpha _1 l)-1\right) \left(\exp(2 \alpha _2
      l)+1\right)\right)\times\\
      &\hspace{1cm}
      \left(a_{11} \left(\alpha _2^2 \left(A_1-\rho_o \right)+B_1\right)+c_{11} \left(\alpha _2^2
      A_2+B_2\right)\right).
\end{aligned}
\end{equation}

Having found the radial displacement $w$ in $[-l,+l]$, we revert to equation \eqref{a1} to find the axial displacement $a_{1}$. A simply calculation yields:
\begin{equation}\label{auno}
a_{1}(x_1)=t\big({\widetilde C}_{1}(x_1)+C_1x_1\big),
\end{equation}
where
\begin{equation}
\begin{aligned}
{\widetilde C}_{1}(x_1):=&A_1\left(\alpha_1k_1\big(\exp(\alpha_1x_1)-\exp(-\alpha_1x_1)\big)\right.\\
&\left.+\alpha_2k_2\big(\exp(\alpha_2x_1)-\exp(-\alpha_2x_1)\big) \right)+\\
&+\frac{B_1}{\alpha_1\alpha_2}\exp\big(-(\alpha_1+\alpha_2)x_1\big)\times\\
&\times\Big(\alpha_1k_2\exp(\alpha_1x_1)\big(\exp(2\alpha_2x_1) -1\big)+\\
&+\alpha_2k_1\exp(\alpha_2x_1)\big(\exp(2\alpha_1x_1) -1\big)  \Big).
\end{aligned}
\end{equation}
The one task remaining is to find $a_{2}$. This we do by integrating \eqref{Ga2}:
\begin{equation}\label{adue}
a_{2}(x_1)=t\big({\widetilde C}_{2}(x_1)+C_2x_1\big),
\end{equation}
where
\begin{equation}
\begin{aligned}
{\widetilde C}_{2}=&A_2\left(\alpha_1k_1\big(\exp(\alpha_1x_1)-\exp(-\alpha_1x_1)\big)+\right.\\
&\left.\alpha_2k_2\big(\exp(\alpha_2x_1)-\exp(-\alpha_2x_1)\big) \right)+\\
&+\frac{B_2}{\alpha_1\alpha_2}\exp\big(-(\alpha_1+\alpha_2)x_1\big)\Big(\alpha_1k_2\exp(\alpha_1x_1)\times\\
&\times\big(\exp(2\alpha_2x_1) -1\big)+\\
&+\alpha_2k_1\exp(\alpha_2x_1)\big(\exp(2\alpha_1x_1) -1\big)  \Big).
\end{aligned}
\end{equation}

\begin{figure}[h]
\centering
\includegraphics[scale=0.6]{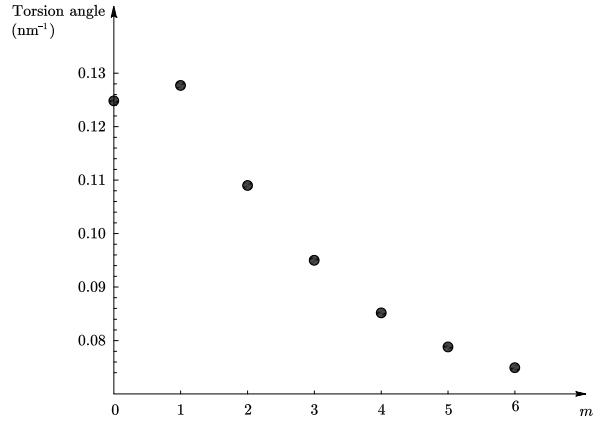}
\caption{Torsion angle for $(6,m)$-CNSs.} \label{torsang}
\end{figure}

\begin{figure}[h]
\centering
\includegraphics[scale=0.6]{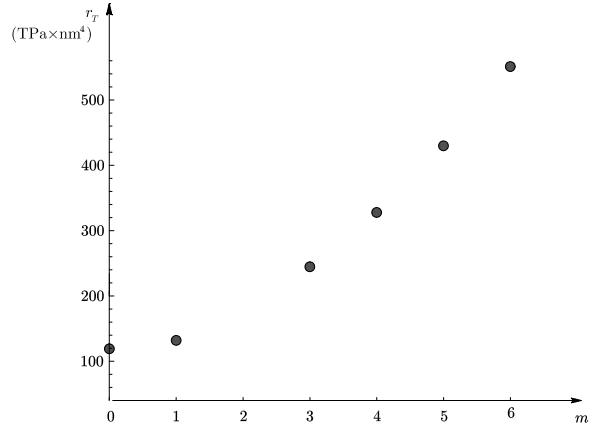}
\caption{Torsion stiffness of  $(6,m)$-CNSs.} \label{rigtor}
\end{figure}

\begin{figure}[h]
\centering
\includegraphics[scale=0.6]{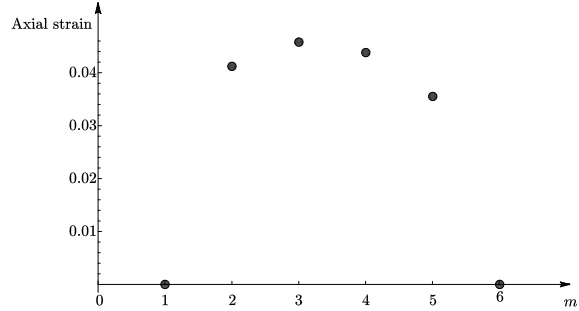}
\caption{Axial strain of twisted $(6,m)$-CNSs.} \label{axstr}
\end{figure}

\remark Needless to say, it is implicit in \eqref{auno} and \eqref{adue} that the parity conditions $a_{\alpha}(0)=0$ $(\alpha=1,2)$ are satisfied. Moreover, as taking a few numerical soundings shows, both \eqref{auno} and \eqref{adue} can be safely replaced by their approximate versions
\begin{equation}\label{aalfa}
 a_{\alpha}(x_1)\simeq t\,C_2x_1\;\;(\alpha=1,2).
\end{equation}
Accordingly, the \emph{torsion angle} and the \emph{torsion stiffness} of a CNS can be evaluated with very good approximation as, respectively,
\[
a_T= a_2^\prime /\rho_o=t\,C_2/\rho_o\quad\textrm{and}\quad s_T=T/a_T=2\pi\rho_o^3/C_2;
\]
Fig.s \ref{torsang} and \ref{rigtor} permit to visualize the chirality dependence inherited by these parameters through $C_2$ and $\rho_o$.\footnote{The computations behind these figures and Fig. \ref{axstr} below have been performed for the following values of the constitutive parameters and the effective thickness, all taken from \cite{BFPPG}:  $E_1=0.784$ TPa, $E_2=0.832$ TPa, $\nu_{12}=0.242$, $\nu_{21}=0.260$, $G=0.424$ TPa; $\varepsilon=0.194$ nm. Moreover, the CNSs in question had all the same slenderness $\rho_o/l=0.25$, and where subject to a circumferential rim load $t=0.1$ N/m.}
Likewise, the axial strain accompanying the twist induced by a given torque $T$ can be evaluated as
$
a_1^\prime=t\,C_1,
$
and depends on chirality through $C_1$ does, as exemplified in Fig. \ref{axstr}.\footnote{This parameter should not be confused with the inverse of the extensional stiffness, that is, the ratio of the applied axial load to the consequent axial strain: to evaluate the latter, it would be necessary to solve the extension problem for CNSs of arbitrary chirality, a doable but cumbersome thing that we defer to another occasion, referring the reader to \cite{Fa2} for the solution of that problem in the case of  zigzag and armchair CNSs.}

\remark
When $\psi=0$, the constants $c_i$ in equation \eqref{eqdiff} become:
\begin{equation}
\begin{aligned}
&c_1=-\frac{2}{3}\varepsilon^3\left(1-\frac{1}{3}\frac{\varepsilon^2}{\rho_o^2}\right)\frac{E_1}{\Delta},\\
&c_2=-\frac{4}{3}\frac{\varepsilon^3}{\rho_o^2}\frac{E_1}{\Delta}\nu_{21},\\
&c_3=-2\frac{\varepsilon}{\rho_o^2}\frac{E_2}{\Delta}\left(\frac{1}{2\frac{\varepsilon}{\rho_o}}\log\frac{1+\frac{\varepsilon}{\rho_o}}{1-\frac{\varepsilon}{\rho_o}}-\nu_{12}\nu_{21}\right),\\
&c_4=0;
\end{aligned}
\end{equation}
when $\psi=\pi/2$, 
\begin{equation}
\begin{aligned}
&c_1=-\frac{2}{3}\varepsilon^3\left(1-\frac{1}{3}\frac{\varepsilon^2}{\rho_o^2}\right)\frac{E_2}{\Delta},\\
&c_2=-\frac{4}{3}\frac{\varepsilon^3}{\rho_o^2}\frac{E_2}{\Delta}\nu_{12},\\
&c_3=-2\frac{\varepsilon}{\rho_o^2}\frac{E_2}{\Delta}\left(\frac{1}{2\frac{\varepsilon}{\rho_o}}\log\frac{1+\frac{\varepsilon}{\rho_o}}{1-\frac{\varepsilon}{\rho_o}}-\nu_{12}\nu_{21}\right),\\
&c_4=0.
\end{aligned}
\end{equation}
Thus, in the case of the \emph{orthotropic} CNSs associated to achiral CNTs, the differential equation \eqref{eqdiff} is homogeneous and the boundary conditions allow to conclude that $w(x_1)\equiv 0$ in $[-l,+l]$; as a consequence, $a_{1}\equiv 0$\,: torsion does not anymore induces axial strain (Fig. \ref{axstr}). 

Coupling of torsional and extensional deformations is the rule with the \emph{generically anisotropic} CNSs associated to chiral CNTs. However, as shown in \cite{BFPPG} in the case of zigzag and armchair CNTs, whose orthotropic anisotropy vanishes for $\rho_0\rightarrow\infty$, torsion/extension coupling must disappear in the same limit for chiral CNTs as well, as a consequence of \eqref{rotko}, where $\Co$, and hence $\widetilde\Co$, becomes isotropic when $\rho_0\rightarrow\infty$.

\section{Conclusions}
We have presented a continuum theory of single-wall CNTs of arbitrary chirality, modeled as anisotropic linearly elastic shells.  Within our theory, a number of equilibrium problems in terms of displacements can be solved explicitly in closed form. As an example, we have worked out in detail, and exemplified in some typical cases, the solution to the `soft' torsion problem, that is, the problem of a CNT subject to end torques.  

There is no need to stress the importance of CNTs as components of a number of NanoElectroMechanical Systems, real and imagined; suffice it to quote a recent paper \cite{ZL}, where NEMS incorporating CNTs as torsional springs, such as nanoscale resonators, are mentioned, and where various chirality-dependent secondary effects, including axial strain, are studied with no size limitations by using molecular dynamics simulations of `hard' torsion problems, that is, problems where the ends of a CNT are given a prescribed relative twist. 

Our explicit formulae, as well as our computations, account both qualitatively and quantitatively for the expected \emph{coupling of torsional and extensional effects} that chirality brings about. Of course, a price to pay for having explicit solutions is to accept the intrinsic limitations in scope of a linear theory. However, as far as possible, the predictions of our theory are in fairly good qualitative and quantitative agreement with the available simulations and experiments. 

We have  various generalizations of our present theory in sight; those that we expect to require a relatively modest effort are: a linear theory covering dynamical situations, such as free or forced vibrations and axial wave propagation; a physically nonlinear theory, resulting from replacing the orthotropic elasticity tensor acting on the linear strain measure \eqref{strain} by the corresponding St.Venant-Kirchhoff fourth-order tensor acting on the  Green-St.Venant  strain tensor; a linear theory of \emph{multi-wall} CNSs, where adjacent CNTs interact `softly', by way of a van der Waals coupling, or `hardly', because of wall-bridging defects.



%
\onecolumn
\section*{Appendix}
\subsection{The coefficients $c_i$ in equation (30)}
\begin{align}
&c_1:=-\frac{2}{3}a_{11}\varepsilon ^3\left(1-\frac{1}{3}a_{11}\left(a_{11}- a_{12}\frac{ c_{11}}{c_{12}}\frac{\left(1+\frac{1}{3}\frac{\varepsilon^2}{\rho_o^2}\right)^2}{1+\frac{\varepsilon^2}{\rho_o^2}}\right)^{-1}\left(a_{11}-2a_{12}\frac{c_{11}}{c_{12}}\frac{1+\frac{1}{3}\frac{\varepsilon^2}{\rho_o^2}}{1+\frac{\varepsilon^2}{\rho_o^2}}  \right)\frac{\varepsilon^2}{\rho_o^2}\right),\\
&c_2:=-\frac{2}{3}\frac{\varepsilon^3}{\rho_o^2}\left(a_{11}- a_{12}\frac{ c_{11}}{c_{12}}\frac{\left(1+\frac{1}{3}\frac{\varepsilon^2}{\rho_o^2}\right)^2}{1+\frac{\varepsilon^2}{\rho_o^2}}\right)^{-1}\left(a_{11}(b_{11}+a_{22})+c_{12}^{-1} \frac{1+\frac{1}{3}\frac{\varepsilon^2}{\rho_o^2}}{1+\frac{\varepsilon^2}{\rho_o^2}}\left(-a_{11}b_{12}c_{11}+\right.\right.\nonumber\\ 
&\hspace{1cm}\left.\left.-2a_{11}a_{12}c_{22}+4a_{12}^2c_{11}c_{12}^{-1}c_{22}-2a_{12}a_{22}c_{11}\right)\right)-\frac{8}{3}a_{12} \frac{c_{22}}{c_{12}}\frac{1}{1+\frac{\varepsilon^2}{\rho_o^2}}\frac{\varepsilon^3}{\rho_o^2},\\
& c_{3}:=-2b_{22}\frac{\varepsilon}{\rho_o^2}\frac{1}{2\frac{\varepsilon}{\rho_o}}\log\frac{1+\frac{\varepsilon}{\rho_o}}{1-\frac{\varepsilon}{\rho_o}}+2a_{11}^{-1}a_{22}b_{11}\left(a_{11}- a_{12}\frac{ c_{11}}{c_{12}}\frac{\left(1+\frac{1}{3}\frac{\varepsilon^2}{\rho_o^2}\right)^2}{1+\frac{\varepsilon^2}{\rho_o^2}}\right)^{-1}\frac{\varepsilon}{\rho_o^2}+\nonumber\\
&\hspace{1cm}-4\frac{\varepsilon}{\rho_o^2}\frac{1+\frac{1}{3}\frac{\varepsilon^2}{\rho_o^2}}{1+\frac{\varepsilon^2}{\rho_o^2}}\left(a_{11}- a_{12}\frac{ c_{11}}{c_{12}}\frac{\left(1+\frac{1}{3}\frac{\varepsilon^2}{\rho_o^2}\right)^2}{1+\frac{\varepsilon^2}{\rho_o^2}}\right)^{-1}\left(a_{12}b_{11}c_{22}c_{12}^{-1}+\right.\nonumber\\
&\hspace{1cm}\left.-a_{12}b_{12}c_{11}c_{22}c_{12}^{-2}+a_{22}b_{12}c_{11}c_{12}^{-1}+a_{22}b_{12}c_{11}c_{12}^{-1}  \right)+4b_{12}c_{22}c_{12}^{-1}\frac{1}{1+\frac{\varepsilon^2}{\rho_o^2}}\frac{\varepsilon}{\rho_o^2},\\
& c_4=2c_{11}\frac{1+\frac{1}{3}\frac{\varepsilon^2}{\rho_o^2}}{1+\frac{\varepsilon^2}{\rho_o^2}}\frac{1}{\rho_o}\left(a_{11}- a_{12}\frac{ c_{11}}{c_{12}}\frac{\left(1+\frac{1}{3}\frac{\varepsilon^2}{\rho_o^2}\right)^2}{1+\frac{\varepsilon^2}{\rho_o^2}}\right)^{-1}\left(a_{22}c_{12}+2a_{12}c_{11}c_{12}^{-2}\frac{1+\frac{1}{3}\frac{\varepsilon^2}{\rho_o^2}}{1+\frac{\varepsilon^2}{\rho_o^2}}  \right)+\nonumber\\
&\hspace{1cm}-2c_{22}c_{12}^{-1}\frac{1}{\rho_o}\frac{1}{1+\frac{\varepsilon^2}{\rho_o^2}}.
\end{align}
%
\subsection{The coefficients $A_\alpha,B_\alpha,C_\alpha$ in equations (32)}
\begin{align}
&A_1:=\left(a_{11}- \frac{\left(1+\frac{1}{3}\frac{\varepsilon^2}{\rho_o^2}\right)^2}{\left(1+\frac{\varepsilon^2}{\rho_o^2}\right)}a_{12} \right)^{-1}\left(\frac{1}{3}a_{11}-\frac{2}{3}\frac{1+\frac{1}{3}\frac{\varepsilon^2}{\rho_o^2}}{1+\frac{\varepsilon^2}{\rho_o^2}}\frac{c_{11}}{c_{12}}a_{12}\right)\frac{\varepsilon^2}{\rho_o},\\
&B_1:=\left(a_{11}- \frac{\left(1+\frac{1}{3}\frac{\varepsilon^2}{\rho_o^2}\right)^2}{\left(1+\frac{\varepsilon^2}{\rho_o^2}\right)}a_{12} \right)^{-1}\left(\frac{1+\frac{1}{3}\frac{\varepsilon^2}{\rho_o^2}}{1+\frac{\varepsilon^2}{\rho_o^2}}\frac{c_{11}}{c_{12}}b_{12}-b_{11} \right)\frac{1}{\rho_o},\\
&C_1:=-\left(a_{11}- \frac{\left(1+\frac{1}{3}\frac{\varepsilon^2}{\rho_o^2}\right)^2}{\left(1+\frac{\varepsilon^2}{\rho_o^2}\right)}a_{12} \right)^{-1}\frac{1+\frac{1}{3}\frac{\varepsilon^2}{\rho_o^2}}{1+\frac{\varepsilon^2}{\rho_o^2}}\frac{c_{11}}{c_{12}}\frac{1}{2\varepsilon},\\
&A_2:=\left(\varepsilon\left(1+\frac{\varepsilon^2}{\rho_o^2} \right)c_{12}  \right)^{-1}\left(\frac{4}{3}\frac{\varepsilon^3}{\rho_o}a_{12}-2\varepsilon\left(1+\frac{1}{3}\frac{\varepsilon^2}{\rho_o^2}\right)a_{12}A_1 \right),\\
&B_2:=\left(\varepsilon\left(1+\frac{\varepsilon^2}{\rho_o^2} \right)c_{12}  \right)^{-1}\left(- 2\frac{\varepsilon}{\rho_o}b_{12}w-2\varepsilon\left(1+\frac{1}{3}\frac{\varepsilon^2}{\rho_o^2}\right)a_{12}B_1 \right),\\
&C_2:=\left(\varepsilon\left(1+\frac{\varepsilon^2}{\rho_o^2} \right)c_{12}  \right)^{-1}\left(1-2\varepsilon\left(1+\frac{1}{3}\frac{\varepsilon^2}{\rho_o^2}\right)a_{12}C_1 \right).
\end{align}

\section*{Acknowledgements}
We thank Nicola Pugno for some useful discussions. AF gratefully acknowledges the financial support of INdAM--GNFM
(Research Project: ``Modelli di strutture sottili per nano- e bio-materiali'').

\end{document}